\documentclass[lettersize,journal]{IEEEtran}
\usepackage{amsmath,amsfonts}
\usepackage{algorithmic}
\usepackage{algorithm}
\usepackage{array}
\usepackage[caption=false,font=normalsize,labelfont=sf,textfont=sf]{subfig}
\usepackage{textcomp}
\usepackage{stfloats}
\usepackage{url}
\usepackage{verbatim}
\usepackage{graphicx}
\usepackage{cite}
\usepackage{tabularx}
\usepackage{booktabs}
\begin{document}

\title{Enhancing Online Learning by Integrating Biosensors and Multimodal Learning Analytics for Detecting and Predicting Student Behavior: A Review}

\author{Alvaro Becerra, Ruth Cobos, Charles Lang
\thanks{Alvaro Becerra (0009-0003-7793-2682) and Ruth Cobos (0000-0002-3411-3009) are with Department of Computer Science Engineering, Universidad Autónoma de Madrid, Spain (e-mail: alvaro.becerra@uam.es, ruth.cobos@uam.es).

Charles Lang (0000-0002-4298-9481) is with Digital Futures Institute, Teachers College Columbia University, New York, USA (e-mail: charles.lang@tc.columbia.edu)}

This is the accepted manuscript of an article that has been accepted for publication in Behaviour \& Information Technology (Taylor \& Francis). The final published version will be available at https://www.tandfonline.com/journals/tbit20 once released.}

% The paper headers
%\markboth{}

%\IEEEpubid{0000--0000/00\$00.00~\copyright~2021 IEEE}
% Remember, if you use this you must call \IEEEpubidadjcol in the second
% column for its text to clear the IEEEpubid mark.

\maketitle

\begin{abstract} 
In modern online learning, understanding and predicting student behavior is crucial for enhancing engagement and optimizing educational outcomes. This systematic review explores the integration of biosensors and Multimodal Learning Analytics (MmLA) to analyze and predict student behavior during computer-based learning sessions. We examine key challenges, including emotion and attention detection, behavioral analysis, experimental design, and demographic considerations in data collection. Our study highlights the growing role of physiological signals, such as heart rate, brain activity, and eye-tracking, combined with traditional interaction data and self-reports to gain deeper insights into cognitive states and engagement levels. We synthesize findings from 54 key studies, analyzing commonly used methodologies such as advanced machine learning algorithms and multimodal data pre-processing techniques. The review identifies current research trends, limitations, and emerging directions in the field, emphasizing the transformative potential of biosensor-driven adaptive learning systems. Our findings suggest that integrating multimodal data can facilitate personalized learning experiences, real-time feedback, and intelligent educational interventions, ultimately advancing toward a more customized and adaptive online learning experience.

\end{abstract}


\section{Introduction}
The popularity of online courses has grown substantially over the last two decades, especially during the COVID-19 pandemic \cite{sharin2021learning}.However, as online learning has expanded, its associated challenges have also become more evident. In particular, the lack of interaction between instructors and students has been identified as a major issue in multiple studies ~\cite{pardo2019using,iraj2020understanding}. Furthermore, students frequently express feelings of isolation, citing a lack of support and feedback from instructors. These factors contribute to a decline in student motivation, ultimately leading to dropout ~\cite{hone2016exploring,topali2019exploring}.

To address these problems in online courses, including popular formats such as MOOCs (Massive Open Online Courses), several Learning Analytics (LA) tools have been developed. Some of them provide descriptive analytics or dashboards \cite{verbert2014learning, lang2017handbook} and visualizations \cite{martinez2024video}, while others provide predictive analytics \cite{martinez2020achievements, sghir2023recent} and others prescriptive analytics or feedback \cite{cobos2023selfregulated,becerra2024generative, knobbout2020learning, topali2024codesign}. However, recent research also highlights how students’ emotional responses to predictive LA tools can vary significantly, with some students feeling motivated while others experience confusion, scepticism, or distress, underscoring the importance of emotionally adaptive learning environments \cite{joseph2021predictive}. Furthermore, several studies have explored student behavior in online courses, analyzing patterns such as demographics, motivation, and time spent on different types of content  with the goal of understanding their impact on dropout rates and performance in the course \cite{rai2016influencing, kizilcec2017diverse, xu2020student, moreno2018analysing}.

In recent years a new trend has emerged, online learning platforms that analyze student characteristics and implement interventions based on a blend of biometric and behavioral data \cite{baro2018integration, hernandez2019edbb, daza2023edbb}.  In particular, these platforms incorporate biosensors and Multimodal Learning Analytics (MmLA) tools to better understand students' behavior \cite{giannakos2022multimodal}. The end goal is to capture, integrate, and analyze massive quantities of data from multiple sources to provide a more comprehensive understanding of the learning process \cite{sharma2020multimodal}.

\subsection{Multimodal Learning Analytics}

Multimodal Learning Analytics (MmLA) is an approach to understanding and enhancing the learning process by integrating diverse data sources \cite{worsley2018multimodal, ochoa2022multimodal}. The term "multimodal" refers to the use of multiple modes of communication and interaction, such as textual, aural, linguistic, spatial, and visual. In the context of MmLA, this means analyzing various forms of data including eye-tracking, heart rate, digital traces, body language, facial expressions, and more.

MmLA combines elements of computer-supported learning analysis and educational data mining, leveraging advanced data processing and modeling techniques like machine learning and deep learning. It aims to provide a comprehensive view of learning by analyzing these various modes of communication and interaction in educational settings, including verbal, non-verbal, and environmental factors. By capturing and analyzing multiple dimensions of the learning experience across different modes of communication and interaction, MmLA offers new insights into learning processes and ways to optimize learning environments. This multi-faceted approach provides a richer, more holistic view of the learning process than traditional single-mode analyses.

Unlike traditional learning analytics that rely primarily on log data, limited to capturing digital traces of student interaction with learning platforms, MmLA draws on diverse data sources. Log data alone often fail to reflect the emotional, social, and embodied aspects of learning that are essential in real-world contexts. In contrast, multimodal data fusion allows researchers to integrate behavioral, physiological, and contextual signals, generating a much richer and more holistic understanding of learning processes \cite{cukurova2020promise}.

One transformative application of MmLA is in providing personalized feedback to students, where generative AI (GenAI) may play a crucial role. Large Language Models (LLMs) can analyze multimodal data—including textual responses, speech patterns, and physiological signals—to generate adaptive, contextually aware feedback. By automating and streamlining the feedback process, GenAI reduces the cognitive load on educators while maintaining high pedagogical standards. AI-driven feedback systems can identify learning difficulties, offer tailored recommendations, and provide metacognitive insights that promote self-regulated learning \cite{giannakos2024promise}. Additionally, the integration of GenAI into learning analytics dashboards enhances real-time data visualization and interpretation. Dashboards powered by LLMs not only facilitate the identification of learning patterns but also dynamically personalize the information presented to educators and students, enabling more effective decision-making in educational settings \cite{alfredo2024designing, becerra2024generative, yan2024vizchat, becerra2025enhancing}.

A key focus in MmLA research is the role of multimodal datasets and sensor technologies, as highlighted in a bibliometric analysis by Pei et al. (2023) \cite{pei2023academic}. The study identifies “Multimodal Dataset” and “Sensor” as critical keywords, emphasizing the significance of integrating diverse data sources to advance learning analytics. The use of multimodal datasets enables a more nuanced analysis of cognitive and emotional aspects of learning. Furthermore, the increasing adoption of sensing technologies, such as wearables, biosensors, and Kinect-based motion tracking, has allowed for real-time analytics and personalized learning interventions. These advancements reinforce the potential of MmLA to drive more effective, data-informed educational strategies.

\subsection{Biosensors}

 In the context of MmLA, multimodality refers to the integration of data from diverse sources and formats, which do not necessarily rely on biosensors. For example, combining textual responses and system interaction logs constitutes a multimodal approach without involving any biometric data. In this review, however, we focus specifically on MmLA studies that incorporate at least one biometric or physiological data source, such as EEG, heart rate, or eye tracking, alongside other modalities. While biosensors are not a separate field from MmLA, our intention in this review is to analyze how their inclusion within multimodal frameworks enhances the ability to capture affective and cognitive dimensions of learning.
 
Several studies~\cite{giannakos2019multimodal, spikol2018supervised} have demonstrated that employing multiple data acquisition devices can lead to better predictions of student success. \cite{yan2025complexity} illustrates how the integration of multimodal sensor data, including body position tracking, audio recordings, and physiological signals such as heart rate, can effectively uncover complex interaction patterns among students during collaborative learning tasks. By leveraging these heterogeneous data streams, the study identifies behavioral dynamics such as turn-taking, social regulation, and coordination strategies, offering valuable insights into how learners engage and co-regulate in group settings. Supporting this perspective, \cite{peng2021recognition} demonstrates that combining multimodal sensor data, including facial expressions, body posture, and speech features, enables the effective recognition of students’ mental states during discussion-based learning. Their approach not only improves the accuracy of mental state classification but also facilitates real-time adaptive educational support, highlighting the value of sensor-rich environments in capturing complex cognitive and emotional dimensions of learning.

Data fusion is a key aspect in MmLA. As highlighted in \cite{chango2022review}, combining diverse data streams enables a more holistic understanding of student behavior, as single-source models are often limited in their ability to capture the complexity of learning dynamics. Concrete examples of this integration can be found in various studies. For instance, in \cite{chango2021improving}, data from an intelligent tutoring system, interaction zones captured via eye-tracking, and emotions detected through webcam videos are combined to predict students' performance. Similarly, in \cite{becerra2023m2lads}, a dashboard is proposed that synchronizes data from an EEG band, a smartwatch, an eye-tracker, and multiple webcams to analyze students' behavior during an online course session. In \cite{liapis2023ux}, user interaction metrics and data from wearable biosensors are leveraged to assess user experience on a MOOC platform. \cite{becerra2024biometrics, becerra2025ai} explore the use of computer vision techniques to detect distractions in online learning by analyzing students' head posture and physiological signals, proposing both unimodal and multimodal models that detect phone usage events, where the multimodal models,combining pose estimation, brainwave activity, attention levels, heart rate, and mental calmness, demonstrate superior performance. In \cite{papoutsaki2018eye}, the relationship between eye gaze and typing behavior is examined through a dataset that captures multimodal interactions, including webcam-based eye tracking, keyboard input, and cursor activity demonstrating that integrating typing behavior as a secondary signal enhances gaze estimation accuracy.

However, while integrating multiple data sources yields better results, monitoring a large number of students with multiple biosensors is both expensive and challenging. As a result, some research focuses on a single biosensor. For instance, in \cite{sharma2020eye}, eye-tracking data is analyzed to understand student motivation and attention, which in turn helps predict performance outcomes. Similarly, in \cite{navarro2024vaad}, eye-tracking data—specifically fixations and saccades—is examined to assess students' visual attention and predict the tasks they are engaged in. In \cite{daza2024deepface}, a webcam-based approach is used to estimate students' attention levels in online learning environments by integrating facial analysis techniques, including head pose detection, facial action units, and eye-blink frequency, to classify attention levels as high or low.

\subsection{Research Questions}

To provide a comprehensive understanding of the landscape of Multimodal Learning Analytics (MmLA), this systematic review analyzes the current technologies and algorithms employed in computer-based online learning. By focusing on the use of biosensors and multimodal data, the study seeks to explore how these methods are being utilized to detect and predict students' behavior, motivation, and presence of learning during digital learning activities. The review is structured around three core aspects:

\begin{itemize}
\item \textbf{RQ1. How have researchers detected and predicted students' behaviors using MmLA and biosensors in online learning and how have they evaluated the presence of learning or if students were motivated while learning? (see Subsection \ref{ss:objective}).} An important aspect of this study is to explore the current approaches and methodologies used to interpret students’ engagement and cognitive-emotional states using multimodal and physiological data. We examine what aspects of learning are being measured and how these aspects are operationalized through biosensor data in MmLA online learning contexts. This helps us understand the conceptual focus of current research and the strategies employed to quantify and evaluate internal learning states.

\item \textbf{RQ2. How has research captured and processed physiological data from students, evaluated and applied algorithms to these signals, and how has the population been distributed in the studies? (see Subsection \ref{ss:biosensors}).} This question focuses on the technical and demographic aspects of the studies. We examine how biosensor data is collected, preprocessed, and analyzed, and who the participants are in these studies. Understanding the diversity of data processing pipelines and algorithmic choices is essential for evaluating the robustness and replicability of findings, while also identifying potential biases in study samples.

\item \textbf{RQ3. What limitations and directions for future work have been identified in the studies? (see Subsection \ref{ss:future work}).}
A critical aim of this review is to examine the self-reported constraints, open questions, and suggested pathways for advancing MmLA with biosensors in online learning. This question seeks to understand the motivations for improvement and the broader implications of current limitations in the field. Addressing this dimension can help uncover systemic challenges and emerging research priorities.

\end{itemize}

\section{Related Work}
This investigation adds to existing reviews in the area of multimodal learning analytics. The scientific literature includes relevant reviews exploring different aspects of the MmLA field. We found several works focusing on specific devices. In \cite{hu2024multimodal}, the authors analyzed the use of virtual reality devices in educational research, examining the various multimodal interactions, modalities, and domains studied. In \cite{liu2024eeg}, the authors reviewed several articles in the EEG-based multimodal emotion recognition field, focusing on the different databases and techniques used to measure and process physiological signals, as well as how to handle missing values. In \cite{han2023multimodal}, EEG bands and neurofeedback are analyzed, and the possibilities and limitations of applying these in online learning are discussed.

There are also works that analyze how to integrate all the signals measured by different biosensors. In \cite{mu2020multimodal}, the types of multimodal data and learning indicators used in MmLA are analyzed, focusing on how multimodal and biometric data are combined to obtain useful and accurate indicators. Similarly, in \cite{chango2022review}, different techniques for data fusion are reviewed, focusing on when the fusion occurs, the various techniques, and the differences in data from online learning and classroom environments.  In \cite{liu2024academic}, the use of biosensors for detecting stress in academic environments is analyzed, including electrocardiogram, electroencephalogram, galvanic skin response, blood pressure, skin temperature, photoplethysmography, and blood volume pulse.

Other reviews have also started to emphasize the theoretical foundations of MmLA. For example, in \cite{giannakos2023role}, the authors conducted a semi-systematic literature review to examine how learning theories are incorporated into MMLA research. They found that while multimodal learning analytics has great potential to deepen the understanding of learning processes, particularly in cognitive, affective, and social domains, there is limited engagement with theory. They also emphasize that although MmLA can generate ecologically valid, real-time insights into learning, this potential is underutilized when theoretical grounding is absent. Similarly, in \cite{ouhaichi2023research}, the authors presented a study aimed at identifying research trends, methodologies, and emerging themes within MMLA. Their review emphasizes the predominance of technology-focused approaches, especially those involving AI and machine learning, while highlighting a lack of theory-driven research and limited real-world validation of MMLA systems. More recently, in \cite{mohammadi2025artificial} a systematic literature review was conducted focused specifically on the intersection of artificial intelligence (AI) and MmLA, highlighting both the transformative role AI can play across the MMLA pipeline, particularly in model learning and feature engineering, and the limited integration of AI in theory-driven components of MmLA research. Their review underscores the need for greater theoretical alignment to fully leverage the benefits of AI-enhanced multimodal learning systems.

Some reviews focus on related fields such as Learning Analytics (LA) or Smart Learning Environments (SLE). In \cite{mangaroska2018learning}, the authors studied learning analytics in learning design, focusing on the synergy between both fields and the perspective categories of the articles, such as individual students, groups, or teachers. In \cite{tabuenca2021affordances}, the authors analyzed the technologies used in SLE, identifying the devices used to capture data, such as cameras, computers, or biosensors. They also studied the techniques used to process all the data and the different pedagogical approaches in SLE. In \cite{banihashem2024learning}, the authors conducted a systematic review on LA in online game-based learning, analyzing how different data types, analytical methods, and stakeholders contribute to optimizing learning in higher education. Their study emphasizes the role of assessment, monitoring, and feedback mechanisms, providing a framework for integrating LA into game-based learning environments.

\begin{table*}[htbp]
\small
\centering
\caption{Comparison of Related Literature Reviews in MmLA and related fields}
\label{tab:comparison}
\begin{tabularx}{\linewidth}{|l|X|X|X|}
\hline
\textbf{Reference} & \textbf{Main Focus} & \textbf{Data/Devices} & \textbf{Contribution} \\ \hline
\cite{hu2024multimodal} & Virtual Reality in education research & Virtual Reality, multimodal interactions & Mapping modalities and domains in Virtual Reality \\ \hline
\cite{liu2024eeg} & EEG-based emotion recognition & EEG band & Databases, processing techniques, missing data handling \\ \hline
\cite{han2023multimodal} & EEG neurofeedback in online learning & EEG bands & Possibilities and Limitations of EEG Bands for Feedback Applications \\ \hline
\cite{mu2020multimodal} & Indicators from multimodal data & Digital logs, physiological sensors (EEG, EDA, HR), eye tracking, webcam & Indicator construction linking multimodal data to cognitive, emotional, and behavioral indicators via data fusion \\ \hline
\cite{chango2022review} & Data fusion techniques &Data from online platforms and classroom sensors (e.g., EEG, logs, video, motion) & Classifies fusion by timing (early, late), methods (e.g., Machine Learning, statistical), and learning contexts (individual vs collaborative) \\ \hline
\cite{liu2024academic} & Stress detection in academia & ECG, EEG, GSR, PPG, etc. & Device comparison and stress-related insights \\ \hline
\cite{mangaroska2018learning} & Learning analytics and learning design & LA tools & Perspective-based taxonomy (student/group/teacher) \\ \hline
\cite{tabuenca2021affordances} & Smart learning environments (SLE) & Cameras, biosensors, computers & Processing techniques and pedagogy analysis \\ \hline
\cite{banihashem2024learning} & LA in online game-based learning & Game data, logs, sensors & Stakeholder focus and feedback mechanisms \\ \hline
\cite{giannakos2023role} & Integration of learning theory in MmLA research & Broad range of multimodal learning data & Identifies a lack of theoretical grounding and offers a framework to align data practices with learning theories \\ \hline
\cite{ouhaichi2023research} & MmLA research trends & Data from multimodal sources (e.g. sensors, speech, body-tracking, Virtual Reality) & Tech-centric trends in MmLA \\ \hline
\cite{mohammadi2025artificial} & AI in MmLA research & Multimodal data & AI techniques for modeling learner behavior, emotion, and cognition \\ \hline
\textbf{This Review} & Biosensor and AI integration for behavior detection and prediction in online contexts & Multimodal data from biosensors (e.g., EEG, EDA, eye-tracking) and online platforms & Comprehensive synthesis of approaches, methods, sensors limitations, and future directions of MmLA applied to online learning\\ \hline
\end{tabularx}
\end{table*}

As summarized in Table~\ref{tab:comparison}, prior literature reviews have addressed specific aspects of the MmLA field, such as the use of particular biosensors, data fusion techniques, or the role of theory and AI. Building upon these efforts, our review offers a complementary and integrative perspective by focusing specifically on the context of computer-based online learning. It jointly analyzes how multimodal physiological signals are captured, processed, and used to detect and predict student behavior in these settings. In addition, it considers how these approaches align with learning goals, their scalability, and the current challenges in real-world validation.

\section{Methodology}\label{apendice:metodos}
The methodology of this study follows three steps: search, selection, and analysis. Furthermore, the process carried out in this review adheres to the PRISMA 2020 protocol \cite{page2021updating} for literature reviews. For the search step, we queried three databases: Scopus, Web of Science (WoS), and the IEEE Computer Society Digital Library (CSDL). Searches were not restricted by publication year. In each database, searches were conducted using a predefined set of terms and were limited to titles, abstracts, and author keywords (see Subsection \ref{ss:search_terms}). For the selection step, we defined inclusion and exclusion criteria (see Subsection \ref{ss:criteria}). Finally, for the analysis step, we followed a selection process and extracted data from the final papers to address the three main research questions (see Subsection \ref{ss:selection}).

\subsection{Search terms}\label{ss:search_terms}
To perform the searches on the databases, we divided the query into four categories: behavior detection, physiological data, learning context, and prediction field. In Table \ref{tab:keywords}, we display all search terms grouped accordingly. These categories are linked using AND and OR operators to ensure the inclusion of at least one term from each category.

\begin{table*}[h]
    \centering
\caption{Terms used in search grouped by category}
    \begin{tabularx}{\textwidth}{|p{2.5cm}|X|}
        \hline
        \textbf{Category} & \textbf{Search Terms} \\
        \hline
        \textit{Behavior Detection} & "behaviour detection", "behavior detection", "behavioral detection", "user behavior", "user behaviour", "activity recognition", "behavior analytics", "behaviour analytics", "multimodal learning analytics", "learning analytics", "multimodal learning", "Biometric* and Behavior", "Biometric* and Behaviour", "multimodal data" \\
        \hline
        \textit{Physiological Data} & "sensor*", "biometric*", "biosensor*", "biosignal*", "physiological data", "biological sensor data", "bioelectric signal*","heart rate", "electrocardiogra*", "ECG", "cardiovascular activit*", "electroencephalography", "electrodermal activit*", "EDA", "GSR", "SCR", "electromyography", "EEG", "blood pressure", "eye movement", "eye-tracking", "eye tracking", "eye tracker", "eye-tracker", "eye ratio", "visual attention", "attention score", "pupil detection", "face descriptor*", "head position", "pose detection", "NIR camera", "infrared camera", "smart watch", "smartwatch", "skin temperature", "inertial sensor", "blood oxygen saturation", "galvanic skin response", "skin conductance response", "electrodermal activity", "multimodal","multimodal data", "multi channel data" \\
        \hline
        \textit{Learning Context} & "e-learning", "elearning", "online learning", "MOOC*", "massive open online course*", "online course*", "learning environment*" \\
        \hline
        \textit{Prediction Field} & "prediction*", "classification*", "Predictive Analysis", "Predictive Analytic*", "machine learning", "algorithm*", "tutoring*", "Prescriptive Analytic*" \\
        \hline
    \end{tabularx}
        
        \label{tab:keywords}
\end{table*}

\subsection{Inclusion and Exclusion Criteria}\label{ss:criteria}
To determine if the results of the database searches were relevant to this study, we established the inclusion and exclusion criteria stated in Table \ref{tab:inclusion}.

\begin{table}[h]
    \centering
    \caption{Inclusion and exclusion criteria}
    \begin{tabularx}{\linewidth}{|X|X|}
    \hline
         \textbf{Inclusion Criteria} & \textbf{Exclusion Criteria}  \\ \hline
         Journal, book and conference articles & Slides, posters and conferences review \\ \hline
         Articles written in English & Articles not written in English or full-text paper not available \\ \hline
         Studies in the field of online learning, MOOCs, remote learning & Studies in the field of collaborative education or face-to-face. Studies in other fields such as medical field or general human activity recognition \\ \hline
         Studies with multimodal data & Studies with no multimodal data \\ \hline
    \end{tabularx}
    
    \label{tab:inclusion}
\end{table}

\subsection{Selection Process and Data Extraction}\label{ss:selection}
\begin{figure}[h]
    \centering
    \vspace{15pt}
    \includegraphics[width=\linewidth]{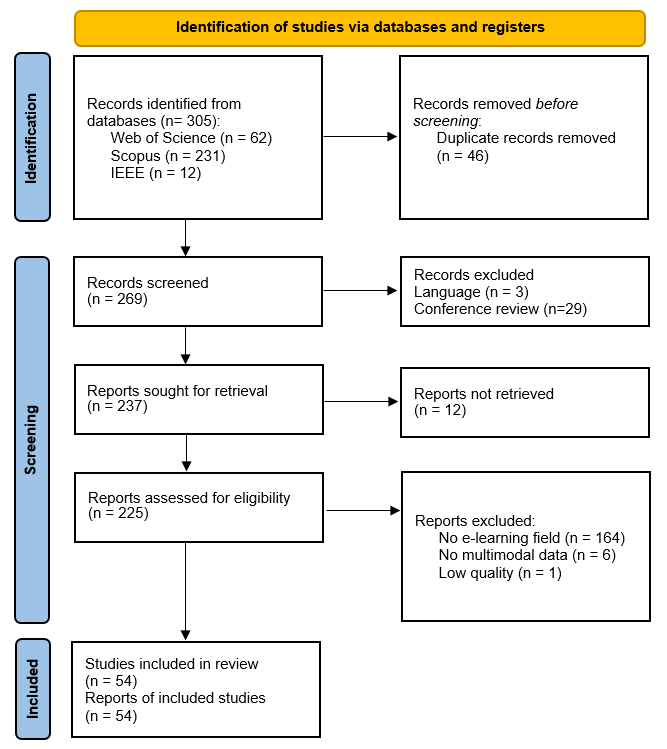}
    \caption{PRISMA flow diagram}
    \vspace{15pt}
    \label{fig:prisma}
\end{figure}

The studies, obtained from the searches carried out in April 2024, in the different 
databases were revised in a two-step selection process. Initially, the title, abstract and keywords of each study were scanned to check if they met the inclusion and exclusion criteria. Subsequently, for all remaining studies, their full texts were retrieved and read in detail. At this stage, studies that did not meet the inclusion/exclusion criteria were excluded. Figure \ref{fig:prisma} illustrates the entire process, which started with 305 studies. We excluded 46 duplicate publications, resulting in 289 unique studies. Following the two-step selection process outlined above, only 54 studies met the inclusion/exclusion criteria.

\begin{figure}[h]
    \centering
    \includegraphics[width=\linewidth]{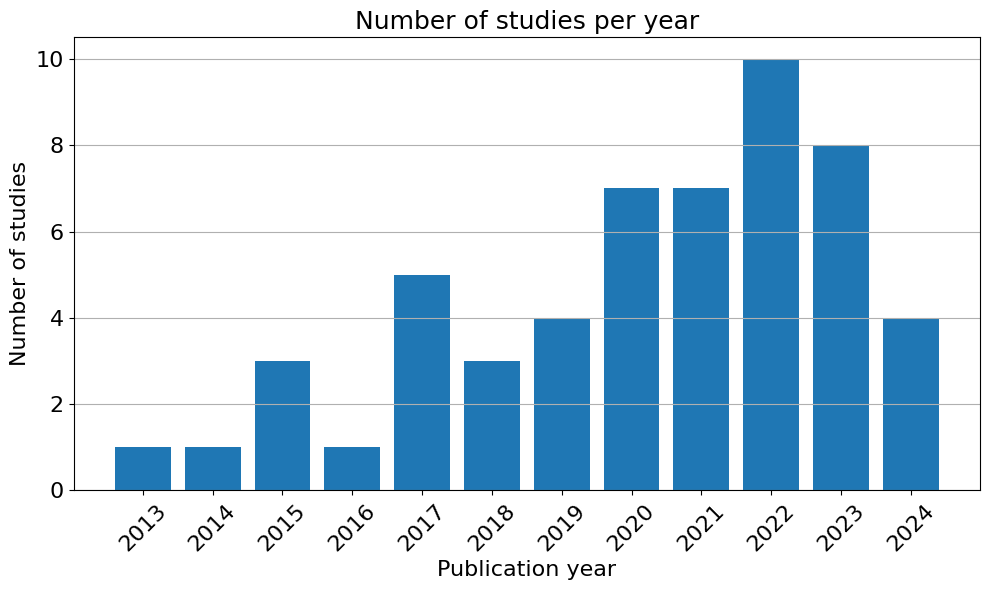}
    \caption{Number of studies found per year of publication}
    \label{fig:year}
\end{figure}

For all selected studies, a set of data was extracted for analysis:

\begin{itemize}
\item General data: Year of publication, authors' countries, publication type, and study objectives.
\item Dataset: Number and gender of participants in the study experiments, average age, other relevant demographic data, employed sensors, and measured variables.
\item Methodology: Pre-processing techniques, algorithms used, prediction targets, and evaluation metrics.
\end{itemize}
The distribution of publication years, as shown in Figure \ref{fig:year}, indicates an increase in interest in studying these MmLA topics for understand and predict students' behavior. Regarding authors' countries, we found authors from 26 countries. However, most of the studies included at least one author from China (13 studies), followed by the United States (12 studies), Japan (7 studies), Germany (5 studies), the Netherlands (5 studies), and Spain (3 studies).

Among the studies, we have 28 conference proceedings (51.85\%), 25 journal articles (46.29\%) and 1 book chapter (1.85\%). Additionally, we identified 9 review papers among them.

\begin{table*}[t]
    \centering
    \caption{Approaches to detect, understand, and predict students' behavior using MmLA and biosensors}
    \begin{tabularx}{\textwidth}{XXX}
    \\ \hline
        \textbf{Ref.} & \textbf{Approaches} & \textbf{Percentage} \\ \hline
        \cite{njeru2017learning}, \cite{coltey2021generalized},\cite{toala2019human}, \cite{munshi2019personalization},\cite{huang2017research}, \cite{wiedbusch2021theoretical},\cite{koch2015architecture}, \cite{azevedo2022lessons},\cite{cloude2022system}, \cite{lester2018metamentor},\cite{caballe2015towards}, \cite{bradavc2022design},\cite{ciolacu2021education}, \cite{duraes2019intelligent},\cite{pham2018predicting} & Presenting a practical approach using MmLA and biosensors & 15 studies, 27.78\% \\ \hline
        \cite{zhou2024state}, \cite{gupta2024artificial},\cite{kawamura2020estimation}, \cite{minematsu2019region},\cite{alyuz2017unobtrusive}, \cite{di2017learning},\cite{di2016learning}, \cite{zhu2023integrating},\cite{liao2022deploying}, \cite{gao2022learning},\cite{minematsu2020visualization}, \cite{sharma2020eye},\cite{sandhu2017evaluation} & Predicting target variables & 13 studies, 24.07\%
        \\ \hline
        \cite{chen2022decade}, \cite{wang2021towards},\cite{praharaj2021literature}, \cite{nandi2020survey},\cite{blikstein2013multimodal},\cite{littlejohn2022professional}, \cite{liu2024eeg}, \cite{han2023multimodal},\cite{mu2020multimodal} & Reviewing studies & 9 studies, 16.67\% \\ \hline
        \cite{kawamura2021detecting}, \cite{li2014online},\cite{abdelmaboud2022machine}, \cite{oder2022automatically},\cite{jia2021design}, \cite{zhang2023evaluation},\cite{abe2015towards} & Detecting students' behavior & 6 studies, 11.11\%
        \\ \hline
        \cite{song2019interest}, \cite{mu2023bcrnet},\cite{li2024measuring}, \cite{lee2023attention} & Classifying students' types & 4 studies, 7.41\%
        \\ \hline
        
\cite{chejara2023exploring}, \cite{liu2023evaluation},\cite{van2020facilitating}, \cite{han2023making},\cite{nguyen2021multimodal}, \cite{montebello2018assisting} & Other approaches & 6 studies, 11.11\%
\\ \hline
    \end{tabularx}

\vspace{15pt}
    \label{tab:objetivos}
\end{table*}

\section{RQ1. Detecting, understanding, and predicting students' behavior}\label{ss:objective}
The approaches of the considered studies to detect, understand, and predict students' behavior using MmLA and biosensors can be categorized into six main groups (Table \ref{tab:objetivos}).

\subsection{Practical Approach Studies}
Among the studies presenting a practical approach using multimodal data and biosensors, we found 6 papers (42.86\%) presenting an intelligent tutoring system that captures data to help students with feedback \cite{njeru2017learning, toala2019human, azevedo2022lessons, ciolacu2021education, duraes2019intelligent, pham2018predicting}. For instance, in \cite{azevedo2022lessons}, an intelligent tutoring system integrates multimodal data, including log files, eye-tracking, facial expressions, and screen recordings, to scaffold self-regulated learning (SRL). The system employs pedagogical agents to provide adaptive feedback, guiding students in setting goals, monitoring their learning, and employing effective cognitive and metacognitive strategies. Another notable example leveraging multimodal data is presented in \cite{pham2018predicting}, where a mobile-based intelligent tutor is designed to predict students' emotions in MOOCs. The system utilizes a dual-camera sensing approach, where the front camera tracks facial expressions while the back camera captures photoplethysmography (PPG) signals to infer heart rate variability. By analyzing these multimodal inputs, the system reliably detects six key emotions and could enable adaptive review mechanisms and targeted interventions. Finally, \cite{duraes2019intelligent} proposes an intelligent tutoring system that leverages biometric behavioral analysis to monitor students’ attention levels in online learning environments. Their system integrates machine learning classifiers to predict user engagement based on keystroke dynamics, mouse interactions, and attention-related performance metrics. By identifying fluctuations in user focus, the system can dynamically adjust the learning experience, providing tailored interventions to enhance learning outcomes.

Additionally, two studies (14.29\%) focus on the development of online learning platforms, introducing innovative architectural frameworks and adaptive learning methodologies \cite{koch2015architecture, caballe2015towards}.

For instance, \cite{caballe2015towards} explores the integration of multimodal data and affective computing into online learning environments. The proposed approach leverages advanced sensors, such as eye-tracking devices and facial expression analysis, along with adaptive interfaces to detect students’ emotions in real time. To provide personalized feedback, the system employs sentiment analysis from textual inputs and multimodal interaction data, adjusting instructional strategies accordingly.

Similarly, \cite{koch2015architecture} proposes an architecture that optimizes real-time feedback by processing multimodal data sources, including video, audio, and interaction logs. This model assesses student engagement and cognitive load through facial expression recognition, voice tone analysis, and keystroke dynamics. Based on these insights, the system delivers adaptive feedback via personalized notifications, real-time content adjustments, and interactive prompts, ensuring students receive timely guidance tailored to their needs.

Another two studies (14.29\%) concentrate on the development of self-regulated learning systems \cite{cloude2022system, lester2018metamentor}.
For example, \cite{cloude2022system} developed a system based on multimodal trace analysis to model SRL in real time. Their approach collects data from multiple sources, including eye movements, concurrent verbalizations, online behavioral logs, facial expressions, and physiological signals. These data sources enable the detection of key moments when students regulate their learning and assess how these strategies impact their performance. Through real-time data processing and analysis, the system identifies when a student demonstrates effective regulation or struggles with learning processes, allowing educators to adjust their intervention strategies accordingly.
 
Other proposed approaches include  a virtual reality system for adaptive immersive learning, which dynamically adjusts content based on user performance and multimodal data, such as eye tracking and physiological responses \cite{coltey2021generalized}, the development of data-driven student tracking system using multimodal data (system logs, eye tracking, and affective states) to personalize scaffolds through conversational agent feedback \cite{munshi2019personalization}, the creation of an individualized student model based on context-awareness, which utilizes contextual data such as location, time or environment information to construct a dynamic, real-time student profile \cite{huang2017research}, development of an adaptive online course  to create personalized study plans, adjusting learning materials based on students’ preferences and ongoing performance assessments\cite{bradavc2022design}, and the implementation of instructors' dashboards which visualizes students' SRL processes using multimodal data to support real-time instructional decision-making and enhance adaptive teaching strategies \cite{wiedbusch2021theoretical}.

\subsection{Prediction-Focused Studies}

Among the studies making predictions, we found that the target variable is students' attention in 4 papers (30.77\%) \cite{gupta2024artificial, kawamura2020estimation, alyuz2017unobtrusive, gao2022learning}. These studies leverage multimodal biosensors, including electroencephalography bands (EEG), photoplethysmography (PPG), electrodermal activity (EDA), and body movement sensors, to capture physiological and behavioral indicators of attention. Various algorithms, including machine learning classifiers such as Support Vector Machines (SVM), Decision Trees, and deep learning models, process these multimodal signals to estimate students' attention levels.

In another 4 papers (30.77\%) the target variable is performance \cite{di2017learning, di2016learning, liao2022deploying, sharma2020eye}. For instance, eye-tracking data has been used to assess student attention and engagement, showing correlations between gaze behavior and academic performance \cite{sharma2020eye}. Similarly, biosensor readings such as heart rate and step count along with environmental data and self-reported performance indicators (stress, productivity, challenge, and abilities) have been utilized to predict learning outcomes \cite{di2016learning, di2017learning}.

Other target variables include predicting whether some reading pages are difficult for students using eye-tracker data \cite{minematsu2019region, minematsu2020visualization}, students' answers using facial features \cite{zhou2024state}, students' activities analyzing the integration of gaze and mouse movement data to detect different digital activities, including literature searching, watching online lectures, reading learning materials, writing assignments or playing games \cite{zhu2023integrating}, and whether a student likes or dislikes a reading using eye-tracker and EEG data \cite{sandhu2017evaluation}.

\subsection{Review Studies}
Among the reviewed studies, two comprehensive reviews specifically focus on the field of Learning Analytics. \cite{chen2022decade} provides a bibliometric analysis based on structural topic modeling of 3,900 Learning Analytics articles published over the past decade, identifying key research themes, trends, and emerging methodologies. Meanwhile, \cite{littlejohn2022professional} examines the role of Learning Analytics in professional education, highlighting its applications in large-scale learning environments, workplace training, and professional development MOOCs. We also found a review focusing on papers using eye-tracker data for measuring and predicting variables such as students' attention or cognitives abilities \cite{wang2021towards}, a review related to different indicators used in collaborative online learning \cite{praharaj2021literature},  a review on emotion recognition using multimodal data, which examines the use of physiological signals (EEG, ECG, EDA), facial expressions, voice, and behavioral data to assess students' emotional states \cite{nandi2020survey}, a review illustrating examples of MmLA techniques applied in education to analyze learning processes, assess student performance, and provide insights into open-ended learning tasks, including programming, engineering design, and hands-on problem-solving \cite{blikstein2013multimodal}, a review on EEG-based emotion recognition, which examines the role of EEG frequency band in identifying emotional states and also discusses the advantages of EEG for emotion recognition, as well as its integration with other physiological signals for improved classification accuracy \cite{liu2024eeg}, a review related to feedback mechanisms for students based on EEG bands (neurofeedback) \cite{han2023multimodal} and a review about how data fusion works in MmLA \cite{mu2020multimodal}.

\subsection{Student Behavior Detection Studies}

Six studies focus on behavioral detection. \cite{kawamura2021detecting} examined wakefulness detection in online learning using multimodal data from heart rate sensors, seat pressure, and facial recognition, identifying awake, drowsy, and asleep states with improved accuracy over traditional log-based methods.\cite{li2014online} investigated user recognition through biometric data collected from mobile sensors, including accelerometers and gyroscopes. 
\cite{abdelmaboud2022machine} focused on detecting non-verbal behavior in e-learning environments leveraged data collection from audio sensors and cameras to classify behavioral cues such as head movements and gaze direction. \cite{oder2022automatically} explored user familiarity while browsing the web using eye-tracking data for distinguishing familiar from unfamiliar users based on fixation duration, scanpaths, and gaze transitions.\cite{jia2021design} analyzed potential cheating behaviors such as unauthorized device usage and multiple individuals in the camera frame.
\cite{zhang2023evaluation} proposed a multimodal data fusion approach for assessing students’ emotional engagement in online education integrating comment data and facial expression analysis. \cite{abe2015towards} investigated activity recognition in online lectures using webcam video data to classify four student activities such as reading text or looking away.

\subsection{Student Classification Studies}

Among the studies focusing on classification, we found two papers classifying students by type of reader using multimodal data such as user interactions, gaze data or engagement metrics\cite{song2019interest, lee2023attention}, one paper that use data from educational exercises (texts and images) to distinguish between different categories of assignments such as writing, mathematics, and logical reasoning \cite{mu2023bcrnet}, and another one classifying students based on measures of their cognitive load in pen-based mobile learning using handwriting, touch gestural, and eye-tracking data\cite{li2024measuring}.

\subsection{Other Approaches}

Finally, among the studies that can't be classified into the previous categories, we found objectives such as identifying indicators for collaborative learning using key behavioral features such as vertical head movements and mouth region activity as potential indicators of collaboration quality \cite{chejara2023exploring}, analyzing the impact of interface design on user engagement by integrating multimodal data, including eye-tracking, EEG, and skin conductance responses \cite{liu2023evaluation}, designing scaffolds to facilitate self-regulated learning \cite{van2020facilitating}, analyzing chatbot interactions in MOOCs, focusing on language markers and user experience \cite{han2023making}, developing and testing a multimodal deep learning model for detecting types of interactions in collaborative learning using data from EDA activity, video, and audio recordings of students working in groups and classifying interactions into metacognitive, socio-emotional, and task execution categories \cite{nguyen2021multimodal}, and studying how ambient factors such as lighting, seating position, and classroom dynamics, impact engagement and learning outcomes \cite{montebello2018assisting}.

\subsection{Predicted Variables}

\begin{figure}[h]
    \centering
    \includegraphics[width=\linewidth]{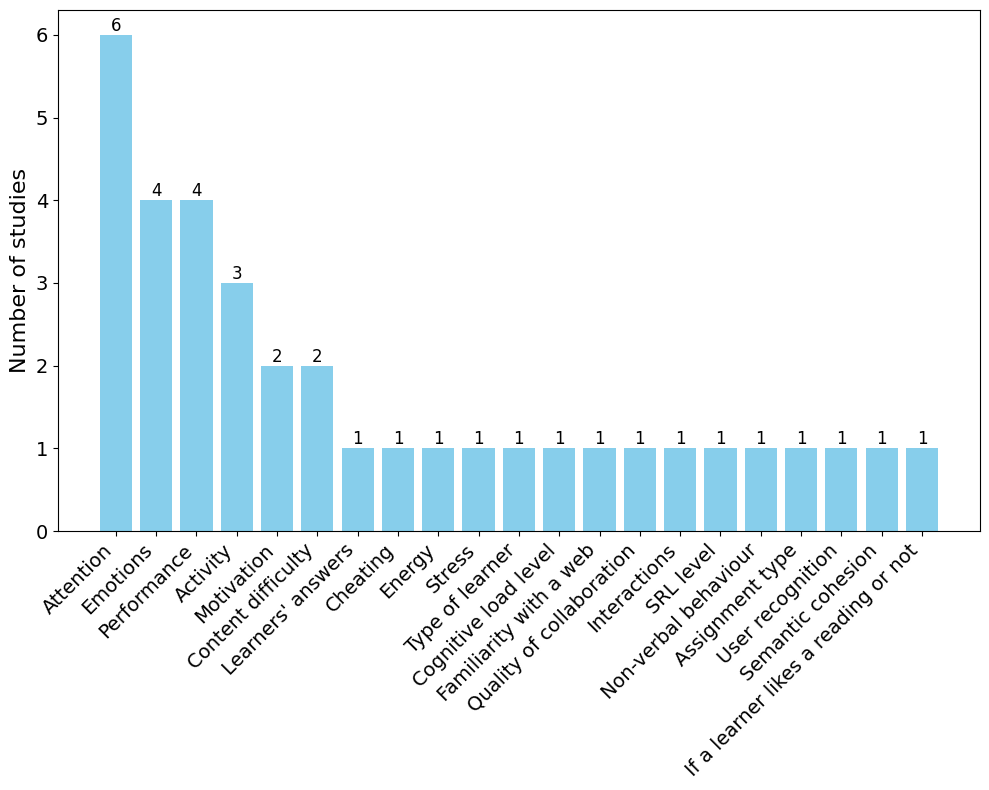}
    \caption{Prediction target in the different studies}
    \label{fig:prediccion}
\end{figure}

In addition, regardless of the purpose of the papers, we found 33 studies that predict variables (Table \ref{tab:prediction}). As shown in Figure \ref{fig:prediccion}, attention is the variable that has been tried to predict the most (6 studies), followed by students' emotions (4 studies) and students' performance (4 studies).

\subsection{Evaluation of Learning, Motivation, and User Perceptions}
Furthermore, 10 studies included evaluations of students' motivation and learning. For checking whether students have learned or not, pretest and posttest is the preferred option (4 studies) \cite{liu2023evaluation,azevedo2022lessons,lee2023attention,sharma2020eye}. Course marks are also used in \cite{liao2022deploying}. For motivation and engagement, the preferred methods are Likert Scales \cite{chejara2023exploring,kawamura2021detecting} and surveys \cite{lee2023attention,sharma2020eye,toala2019human,minematsu2019region}. A set of rules based on the data from an EEG band and an eye-tracker is also used in \cite{sandhu2017evaluation}.

In addition to these measures, only one study also collected students’ subjective perceptions regarding the usability, usefulness, or overall experience with the tools. This study \cite{liu2023evaluation} included semi-structured interviews to gather participants' suggestions about the system’s interface design, such as the layout of navigation and annotation areas, highlight color options, and visibility permissions. These interviews provided valuable qualitative feedback that informed future design improvements of the collaborative reading system they developed. 

In contrast, the authors of \cite{caballe2015towards} propose a clear plan to gather participants’ perceptions of the proposed system through structured questionnaires. They explain that, following the system’s use in pilot settings, participants will be asked to complete a questionnaire designed to assess various dimensions, including the system’s educational value, technical performance, usability, emotional experience during use, and overall satisfaction. According to the authors, the System Usability Scale (SUS) will be employed to evaluate usability, while the Computer Emotion Scale (CES) will be used to assess emotional states. Furthermore, a 5-point Likert scale is planned to measure the perceived value of the system, and open-ended questions will invite participants to provide additional feedback and suggestions for improvement. However, the results of these evaluations are not available.

\begin{table*}
\caption{Summary of prediction target variables and sensors, data sources and signals used}
    \begin{tabularx}{\textwidth}{lXX}
\toprule
                             \textbf{Ref.} & \textbf{Prediction Target Variable}&                     \textbf{Sensors, Data Sources and Signals Used} \\
\midrule
        \cite{abe2015towards} & Activity & webcam \\ \hline
        \cite{duraes2019intelligent} & Activity & eye-tracker, keyboard, logs \\ \hline
        \cite{zhu2023integrating} & Activity & eye-tracker, mouse \\ \hline
        \cite{mu2023bcrnet} & Assignment type & - \\ \hline
        \cite{alyuz2017unobtrusive} & Attention & webcam, mouse \\ \hline
        \cite{gao2022learning} & Attention & EEG, EDA, BVP, SKT, IBI \\ \hline
        \cite{gupta2024artificial} & Attention & EEG, webcam \\ \hline
        \cite{kawamura2020estimation} & Attention & webcam, seat pressure \\ \hline
        \cite{kawamura2021detecting} & Attention & HR, seat pressure, webcam \\ \hline
        \cite{toala2019human} & Attention & mouse, keyboard \\ \hline
        \cite{jia2021design} & Cheating & webcam, microphone \\ \hline
        \cite{li2024measuring} & Cognitive load level & eye-tracker \\ \hline
        \cite{minematsu2019region} & Content difficulty & eye-tracker \\ \hline
        \cite{minematsu2020visualization} & Content difficulty & eye-tracker, mouse \\ \hline
        \cite{caballe2015towards} & Emotions & webcam, microphone, HR, mouse, logs, keyboard \\ \hline
        \cite{liu2024eeg} & Emotions & EEG, EMG, EOG, SKT, GSR, BVP, RSP, ECG, microphone,webcam, eye-tracker \\ \hline
        \cite{pham2018predicting} & Emotions & webcam \\ \hline
        \cite{zhang2023evaluation} & Emotions & webcam, logs \\ \hline
        \cite{oder2022automatically} & Familiarity with a web & eye-tracker \\ \hline
        \cite{nguyen2021multimodal} & Interactions & EDA, webcam, microphone \\ \hline
        \cite{zhou2024state} & students' answers & controller, webcam \\ \hline
        \cite{sandhu2017evaluation} & Motivation, If a student likes a reading or not & EEG, eye-tracker, HRV, GSR \\ \hline
        \cite{abdelmaboud2022machine} & Non-verbal behavior & webcam, microphone \\ \hline
        \cite{di2016learning} & Performance & Not specified \\ \hline
        \cite{di2017learning} & Performance & HR, cell phone \\ \hline
        \cite{liao2022deploying} & Performance & logs, survey \\ \hline
        \cite{sharma2020eye} & Performance, Motivation & eye-tracker, survey \\ \hline
        \cite{chejara2023exploring} & Quality of collaboration & webcam, logs \\ \hline
        \cite{chen2022decade} & Semantic cohesion & - \\ \hline
        \cite{munshi2019personalization} & SRL level & logs, survey \\ \hline
        \cite{ciolacu2021education} & Stress, Energy & HR, HRV, environmental data \\ \hline
        \cite{lee2023attention} & Type of student & webcam, survey \\ \hline
        \cite{li2014online} & User recognition & controller \\
\bottomrule
\end{tabularx}

\label{tab:prediction}
\end{table*}

\section{RQ2. Biometric sensors, algorithms and metrics}\label{ss:biosensors}

\subsection{Biometric Sensors}

\begin{figure}[ht]
    \centering
    \includegraphics[width=\linewidth]{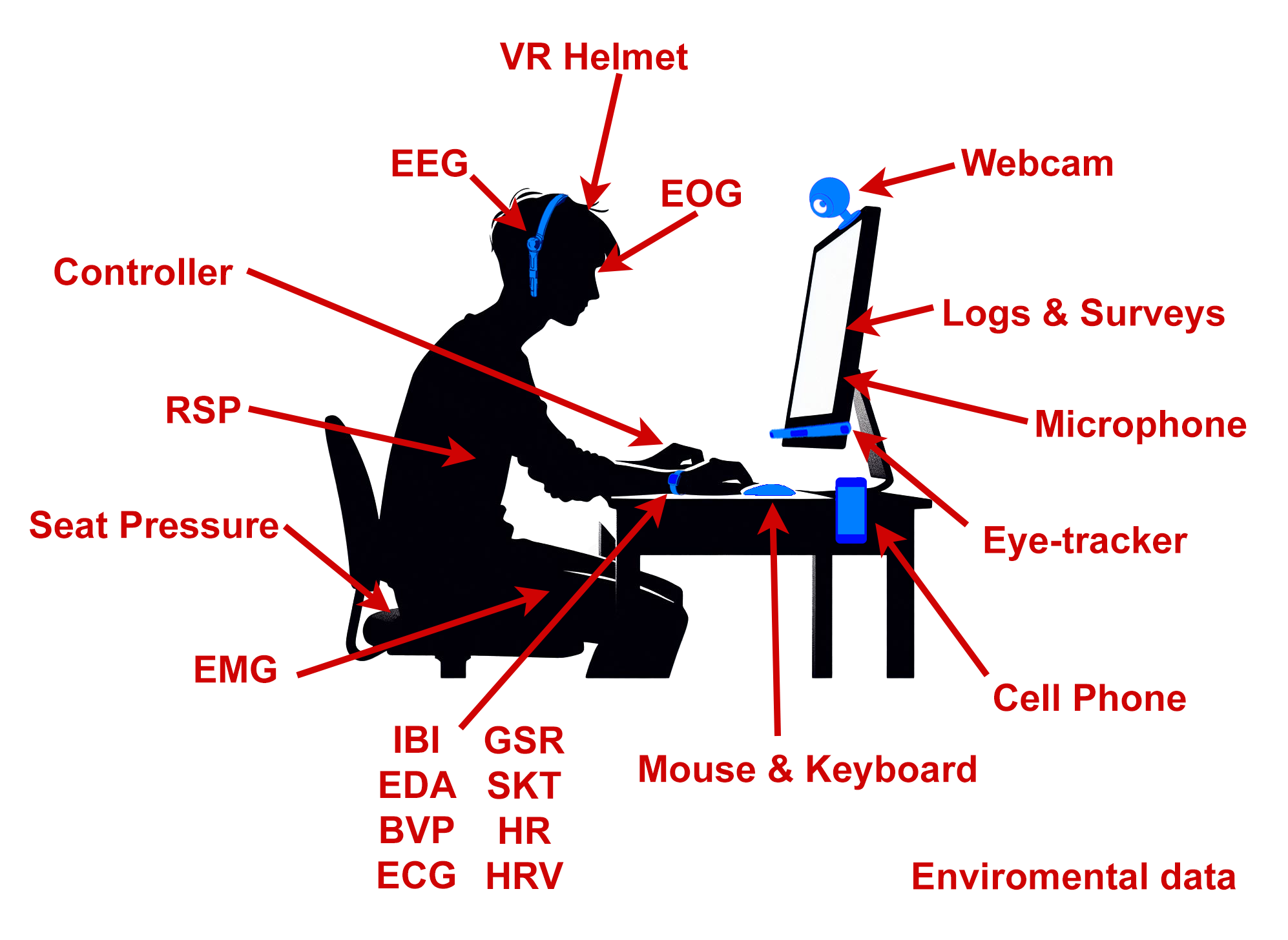}
    \caption{Example of an acquisition setup using all sensors, data sources and signals}
    \label{fig:diagrama}
\end{figure}

We have identified 24 sensors used in the different studies. A diagram illustrating all sensors in an acquisition setup is shown in Figure \ref{fig:diagrama}.

\subsubsection{Webcam}
\begin{figure}[ht]
    \centering
    \includegraphics[width=\linewidth]{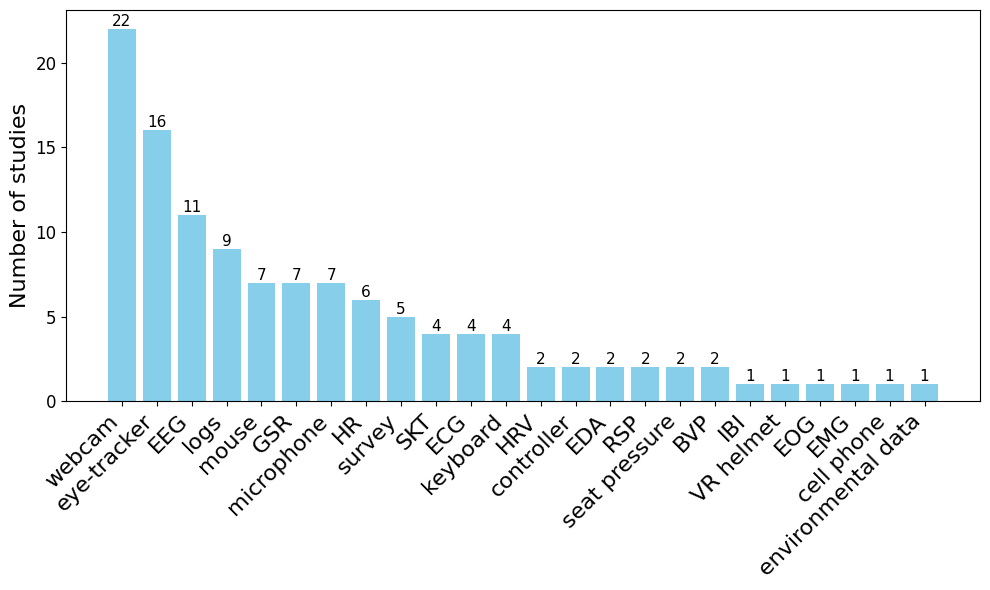}
    \caption{Sensors, signals and data sources used in the different studies}
    \label{fig:sensors}
\end{figure}

As depicted in Figure \ref{fig:sensors}, the most commonly employed sensor is the webcam \cite{zhou2024state, chejara2023exploring, kawamura2021detecting, abdelmaboud2022machine, gupta2024artificial, nguyen2021multimodal, kawamura2020estimation, alyuz2017unobtrusive, azevedo2022lessons, jia2021design, cloude2022system, littlejohn2022professional, lester2018metamentor, montebello2018assisting, caballe2015towards, zhang2023evaluation, liu2024eeg, lee2023attention, pham2018predicting, abe2015towards}. The webcam is a relatively low-cost and widely accessible biosensor that provides a non-intrusive means of capturing valuable behavioral data. Its ability to analyze facial expressions, eye movements, and other physiological cues makes it a powerful tool for assessing posture and gestures, detecting distractions, and evaluating emotional states in various settings. Computer vision and machine learning techniques are employed to detect human emotions through micro-expressions, muscle movements, and other facial features. \cite{canedo2019facial} highlights the effectiveness of these methods in improving the accuracy of emotion recognition. In parallel with facial expression analysis, human pose estimation has emerged as a key technique for interpreting body language and movement patterns. \cite{wang2021deep} provides an extensive review of deep learning approaches for 3D human pose estimation, discussing their applications in fields such as human–computer interaction, video surveillance, and biomechanics. These methods aim to infer 3D joint positions from 2D images or video streams, often under challenging real-world conditions like occlusions and complex backgrounds. Building upon these advances, \cite{zheng2023deep} presents a comprehensive survey of deep learning-based methods for both 2D and 3D human pose estimation. Their work outlines key trends in pose estimation, including the use of monocular RGB inputs, the integration of temporal information from video sequences, and the incorporation of parametric human body models to reconstruct full-body 3D meshes.

These computer vision techniques have been increasingly adopted in MmLA research. For instance, in \cite{azevedo2022lessons, cloude2022system}, the webcam is utilized to capture students’ facial expressions, which are later analyzed to infer their emotional states during the learning process. This emotional data helps understand affective responses and their impact on learning, enabling intelligent tutoring systems to adapt interventions and support SRL more effectively. Webcams can also detect whether students are looking away, signaling distraction, have their eyes closed, indicating fatigue, or remain focused on the screen \cite{abe2015towards}. By analyzing facial expressions and head movements, they provide real-time insights into engagement levels, helping to identify shifts in attention and engagement states. Moreover, webcam-based attention feature extraction has proven to be an effective method for tracking attention regulation behaviors in online learning environments. In \cite{lee2023attention}, behavioral cues such as eye movements, blinks, and facial expressions were used to predict students' attention levels. Similarly, in \cite{kawamura2020estimation}, the authors utilized webcam-based facial analysis to extract blinks and compute eye-opening rates using facial landmarks around the eyelids. Additionally, they analyzed head rotation and translation along the X, Y, and Z axes to detect signs of drowsiness, leveraging these motion-based features to assess attention levels in video-based learning environments.

\subsubsection{Eye-tracker}
The eye tracker is the second most commonly used biosensor \cite{liu2023evaluation, van2020facilitating, song2019interest, njeru2017learning, minematsu2019region, li2024measuring, azevedo2022lessons, oder2022automatically, cloude2022system, lester2018metamentor, liu2024eeg, zhu2023integrating, minematsu2020visualization, sharma2020eye, duraes2019intelligent, sandhu2017evaluation}. Eye trackers enable precise monitoring of gaze direction, fixation duration, saccades, and pupil dilation, providing detailed insights into visual attention patterns, reading behavior, and emotional responses in online learning environments. For example, they help researchers assess information processing efficiency, engagement fluctuations, and decision-making processes. However, while eye trackers offer high-resolution gaze data, they can be more expensive compared to other biosensors, potentially limiting their accessibility in some research and educational contexts. 

Research has shown that pupil dilation systematically increases in response to heightened processing demands during cognitive tasks such as memory recall, language comprehension, reasoning, and attention. In particular, the amplitude of pupil dilation correlates with the difficulty and complexity of the task, making it a sensitive and noninvasive measure of internal cognitive states such as attention, memory load, and information processing effort \cite{beatty1982task}. Pupil dilation has also been widely used as an indicator of sustained attention and mental workload in applied cognitive research. In \cite{vcegovnik2018analysis}, an eye-tracker was employed to continuously measure pupil diameter while participants performed a sustained attention task under varying workload conditions. The results showed that increased cognitive load was significantly associated with larger pupil sizes. More recently, \cite{sankar2025measuring} extended the use of eye-tracking by combining gaze-based and pupil-based metrics to capture two complementary dimensions of cognitive engagement: attention and absorption. In their study, fixation duration was used to measure visual attention, while pupil dilation served as an indicator of absorption, defined as the state of being fully immersed in a task. Their results showed that these measures can track engagement dynamics over time and are especially useful in emotionally demanding settings where traditional flow-based assessments may fall short.

Building on these methodological advancements, the reviewed studies have leveraged eye-tracking data to explore more granular aspects of user interaction and engagement in educational settings. Heatmaps of fixations on a screen, generated using eye-tracking data, have been widely used to gain insights into content difficulty assessment and student engagement levels. In \cite{minematsu2020visualization}, fixation heatmaps were employed to identify specific areas in digital textbooks where students struggled, correlating gaze concentration with perceived difficulty and enabling region-wise difficulty estimation based on fixation density and dwell time. Similarly, in \cite{sharma2020eye}, eye-tracking heatmaps were utilized to analyze attention distribution in MOOCs, focusing on how students interact with instructional videos. By examining variables such as content coverage, scanpath patterns, and students' ability to visually follow the instructor’s deictic gestures and explanations, the authors found that students with higher fixation density on instructional content tended to perform better, whereas lower gaze engagement correlated with reduced learning gains.

Similarly, eye trackers have also been employed in intelligent tutoring systems  to provide complementary insights into students’ cognitive and metacognitive processes. By capturing fixation duration, saccades, and transitions between areas of interest, researchers were able to identify moments where students struggled with instructional material and analyze how frequently students fixated on pedagogical agents, instructional content, and learning prompts to assess engagement levels and responsiveness to system interventions \cite{azevedo2022lessons, cloude2022system}.

\subsubsection{EEG band}
The EEG band is the third most commonly used biosensor \cite{liu2023evaluation, gupta2024artificial, liu2024eeg, han2023multimodal, gao2022learning, sandhu2017evaluation}. EEG bands provide real-time brain activity measurements, allowing researchers to assess mental workload, cognitive load, engagement, and attention levels. EEG signals are typically analyzed through different frequency bands (delta, theta, alpha, beta, and gamma), each associated with distinct cognitive functions relevant to learning. However, EEG bands are generally more intrusive than webcam and eye-trackers.

Despite being more intrusive, EEG remains one of the most effective and objective methods for measuring attention and cognitive states. EEG signals are directly derived from the electrical activity of neurons and provide a real-time, high-temporal-resolution window into brain dynamics. Unlike webcam-based facial recognition or eye-tracking, which rely on indirect behavioral indicators (e.g., gaze patterns, pupil dilation, or facial expressions), EEG captures neurophysiological responses that reflect internal cognitive processes such as learning, decision-making, stress, and mental effort \cite{chen2018effects, li2011real}.

Each EEG frequency band reflects specific cognitive or emotional states. For example, alpha waves (8–13 Hz) are typically associated with relaxed wakefulness and reduced during cognitively demanding tasks \cite{klimesch1993alpha, berger1929elektroenkephalogramm}. Excessive alpha activity may indicate low engagement or difficulty focusing. In contrast, beta waves (13–30 Hz) increase during problem-solving, decision-making, and active concentration, making them reliable markers of cognitive workload \cite{lin2018mental, chen2017assessing}. As \cite{hall2020guyton} notes, beta waves often replace alpha waves when attention shifts toward external tasks or mental processing.

Gamma waves ($>30$ Hz) have been linked to high-level cognitive functions such as working memory, attention, and perception \cite{kaiser2003induced, herrmann2001gamma}. Reduced gamma activity has been associated with learning difficulties and attentional deficits \cite{lee2003synchronous}. Although delta waves ($<4$ Hz) are most prominent during deep sleep, emerging research suggests they also play a role in internal attention and decision-making processes during wakeful states \cite{harmony1996eeg, traub2024neocortical}.

Theta waves (4–8 Hz), traditionally linked to emotional stress or meditative states, have also been observed during tasks that require high memory load or abstract reasoning. In fact, increased theta activity has been recorded in cognitively demanding conditions, while beta activity often decreases, reflecting a shift in cognitive strategies \cite{mecklinger1992event, cahn2006meditation}.

These spectral dynamics are not only relevant for theoretical models of cognition but have also been exploited in real-world monitoring systems. For instance, recent reviews have shown that even low-cost consumer EEG headsets can detect changes in theta and alpha activity related to drowsiness or fatigue, with some systems achieving over 90\% classification accuracy \cite{larocco2020systemic}.

Building upon these neurophysiological insights, the reviewed studies in the field of MmLA have leveraged EEG frequency band characteristics to detect and interpret cognitive and emotional states in educational contexts. In \cite{liu2024eeg}, EEG frequency bands were utilized for emotion recognition, leveraging their strong correlation with central nervous system activity. Each frequency band serves a distinct function in capturing emotional states: the alpha waves (8–13 Hz) is commonly used to assess relaxation and mental workload, the beta waves (13–30 Hz) is employed to detect heightened cognitive engagement and stress, and the gamma waves ($>30$ Hz) is crucial for analyzing emotional perception and neural synchrony. By monitoring these EEG bands, researchers can evaluate students' cognitive engagement, stress levels, and motivation, providing valuable insights for optimizing learning environments

In \cite{gupta2024artificial}, beta waves (12-30 Hz) have been identified as key indicators of alertness and active cognitive engagement, whereas alpha waves (8-12 Hz) are associated with passive attention and relaxation. In that study, the authors predicted attention using machine learning models, such as SVM, based on EEG features. Additionally, they applied clustering techniques, such as K-means, to effectively distinguish between attentive and inattentive states by analyzing the separability of EEG frequency bands.

Similarly, in \cite{gao2022learning}, EEG data were used as ground truth labels to train machine learning models in predicting attention levels based on multimodal physiological signals collected from a wrist-worn device. By correlating EEG-based attentional states with features extracted from blood volume pulse (BVP), inter-beat intervals (IBI), electrodermal activity (EDA), and skin temperature (SKT), researchers aimed to infer attention levels without relying on direct EEG recordings. This approach opens the possibility of eliminating the need for EEG bands in more realistic and naturalistic learning setups outside the controlled laboratory environment. 

\subsubsection{Logs, mouse data and microphone}
Additionally, logs \cite{chejara2023exploring, njeru2017learning, han2023making, munshi2019personalization, cloude2022system, caballe2015towards, zhang2023evaluation, liao2022deploying, duraes2019intelligent}, mouse data \cite{van2020facilitating, song2019interest, toala2019human, alyuz2017unobtrusive, caballe2015towards, zhu2023integrating, minematsu2020visualization}, microphone recordings  \cite{van2020facilitating, abdelmaboud2022machine, nguyen2021multimodal, jia2021design, cloude2022system, caballe2015towards, liu2024eeg} are often used.

On one hand, logs and mouse data are very useful for understanding students' interaction patterns, as they provide detailed insights into behavioral dynamics and response strategies. Logs capture sequential actions, navigation paths, and time spent on tasks, offering a structured view of how students interact with digital content. Mouse movement data, on the other hand, reveals subtle behavioral cues such as hesitation, exploration strategies, and decision-making processes. 

For example, \cite{zhu2023integrating} highlighted how mouse trajectories, click patterns, and movement dynamics correlate with cognitive engagement and different digital activities. By analyzing these patterns along with gaze patterns, they were able to identify learning-related and non-learning-related behaviors.

In \cite{caballe2015towards} authors emphasized how mouse movement trajectories, click frequencies, and scrolling behaviors can serve as indicators of cognitive engagement, frustration, and hesitation during learning tasks. Specifically, they found that frequent erratic mouse movements and rapid clicking patterns were often associated with increased frustration or confusion, while smooth, deliberate movements correlated with focused engagement. Moreover, session logs were used to track the sequence of interactions, task completion times, and user navigation patterns, enabling a more comprehensive analysis of learning behaviors.

Microphones capture both verbal and non-verbal vocal behaviors, such as tone, pitch, speech pauses, and vocal stress. For instance, \cite{abdelmaboud2022machine} utilized microphones to analyze students' non-verbal cues using machine learning models. Their approach involved processing audio recordings of students’ speech and vocal expressions, which were then classified to detect patterns associated with engagement, frustration, or confusion.

One of the most powerful applications of microphone data is speech transcription, which allows researchers to analyze the content, structure, and delivery of students’ spoken interactions. \cite{cloude2022system} highlighted the integration of real-time speech transcriptions with other modalities, such as video, gaze tracking, and behavioral log data, to examine students' SRL processes, as through transcription analysis, researchers can identify patterns in students' verbalized reasoning, self-explanations, and metacognitive reflections.

\subsubsection{Galvanic skin response and heart rate sensors}
Galvanic skin response (GSR) \cite{liu2023evaluation, littlejohn2022professional, liu2024eeg, sandhu2017evaluation} and heart rate (HR) \cite{kawamura2021detecting, cloude2022system, littlejohn2022professional, di2017learning, caballe2015towards} measurements are also frequently utilized. 

Heart rate and galvanic skin response have been extensively studied across various domains, including psychology, neuroscience, human-computer interaction, and affective computing, as reliable indicators of physiological arousal and emotional processing. These signals are commonly used to assess stress, attention, and engagement, as well as to distinguish between different affective states. For instance, in \cite{levenson2003autonomic} both heart rate and galvanic skin response varied significantly across emotions such as anger, fear, and disgust. For example, fear elicited increases in both HR and GSR, while disgust showed a different pattern with decreases in HR and less pronounced skin conductance changes. In addition, increased heart rate is frequently observed in situations involving stress, anxiety, or physical activity, whereas reduced heart rate is generally linked to states of relaxation and calmness \cite{kim2018stress, choi2009using}.

Building on this evidence, GSR, has been widely used in MmLA to assess students’ emotional arousal and engagement during digital learning activities. For instance, in \cite{liu2023evaluation}, they found that GSR peaks reflected variations in excitement levels, suggesting that GSR can effectively capture emotional fluctuations during learning interactions.

Regarding heart rate, \cite{cloude2022system} explored the use of heart rate sensors to track moment-to-moment fluctuations in arousal during SRL tasks. Their study found that higher heart rates were often associated with intense cognitive effort and engagement, whereas lower heart rates corresponded to reduced attention and potential disengagement. Similarly, \cite{kawamura2021detecting} leveraged heart rate sensors to detect attention, revealing that a decline in heart rate was a strong indicator of fatigue and decreased wakefulness.

Other sensors, such as the inter-beat intervals sensor (IBI) \cite{gao2022learning} or electrooculography sensor (EOG) \cite{liu2024eeg}, are less commonly utilized.


\begin{table*}
\vspace{15pt}
\caption{Summary of data sources and signals, number of participants, algorithms, and metrics used for monitoring students and making predictions}
 \begin{tabularx}{\textwidth}{lXXXXX}
\toprule
                             \textbf{Ref.} & \textbf{Data Sources and Signals} &             \textbf{Students Monitored} &                                         \textbf{Algorithms} &  \textbf{Metrics}\\
\midrule
             \cite{zhou2024state} &                                 controller, webcam &                                                 20 & Markov chain Monte Carlo, Item Response Theory Deep Learning &                                  Accuracy,F1-score \\ \hline
      \cite{chejara2023exploring} &                                       webcam, logs &                                                 69 &                                        k-means, RF &                               Spearman correlation \\ \hline
         \cite{liu2023evaluation} &                              GSR, eye-tracker, EEG &                                                 47 &                                                - &                                             T-test \\  \hline
     \cite{kawamura2021detecting} &                          HR, seat pressure, webcam &                                                 48 &                                  SVM, RF, CatBoost &                                           F1-score \\ \hline
              \cite{li2014online} &                                         controller &                                                 28 &                          RF, kNN, SVM, Naive Bayes &                                           Accuracy \\ \hline
             \cite{han2023making} &                                       survey, logs &                                                 42 &                                                - &                                                - \\ \hline
    \cite{abdelmaboud2022machine} &                                 webcam, microphone &                                               Not specified &        Sparse Autoencoder, Ant Colony Optimization &                                       Accuracy,MSE \\ \hline
            \cite{toala2019human} &                                    mouse, keyboard &                                                 14 &                                                - &                                      Attention (\%) \\ \hline
 \cite{munshi2019personalization} &                                       logs, survey &                                                 99 &                                                - &                                                - \\ \hline
       \cite{gupta2024artificial} &                                        EEG, webcam &                                                500 &         Logistic regression, Ridge regression, SVM &              Accuracy, Precision, Recall, F1-score \\ \hline
      \cite{nguyen2021multimodal} &                            EDA, webcam, microphone &                                                 94 &                                                CNN &                                                - \\ \hline
    \cite{kawamura2020estimation} &                              webcam, seat pressure &                                                 49 &                                           Catboost &                                           Macro-F1 \\ \hline
       \cite{minematsu2019region} &                                        eye-tracker &                                                 15 &                                                CNN &                                                - \\ \hline
      \cite{alyuz2017unobtrusive} &                                      webcam, mouse &                                                 17 &                                               Uni-modal classifier &                                           F1-score \\ \hline
           \cite{li2024measuring} &                                        eye-tracker &                                                 19 &                                                SVM &                                      AUC, F1-score \\ \hline
     \cite{oder2022automatically} &                                        eye-tracker &                                                 78 &                           Logistic regression, SVM &              Accuracy, Precision, Recall, F1-score \\ \hline
            \cite{di2017learning} &                                     HR, cell phone &                                                  9 &                         Linear Mixed Effect Models &                                                $R^2$ \\ \hline
       \cite{zhang2023evaluation} &                                       webcam, logs &                                                100 &                       Adaboost Classification, kNN &                           Accuracy, MSE, MAE, MAPE \\ \hline
                \cite{liu2024eeg} & EEG, EMG, EOG, SKT, GSR, BVP, RSP, ECG, microphone,webcam, eye-tracker & 32/23/15/30/56/40 & CNN, Neural Network & Accuracy, Std, Recall, Precision, F1-score, p-value \\ \hline
        \cite{zhu2023integrating} &                                 eye-tracker, mouse &                                                 21 &                                CNN, Neural Network &                                           F1-score \\ \hline
          \cite{lee2023attention} &                                     webcam, survey &                                                 60 & k-means, spectral clustering, AdaBoost, SVM, kNN, RF &                                           Accuracy \\ \hline
         \cite{liao2022deploying} &                                       logs, survey &                                                 51 &   RF, SVM, LSTM, Neural Network, Linear regression &                                          $R^2$, KR20 \\ \hline
           \cite{gao2022learning} &                            EEG, EDA, BVP, SKT, IBI &                                                 18 & SVM, RF, Decision Tree, Naive Bayes, Bayesian Network, Logistic regression, kNN &              Accuracy, Precision, Recall, F1-score \\ \hline
      \cite{ciolacu2021education} &                        HR, HRV, environmental data &                                                Not specified &                                        Rule system &                                                - \\ \hline
\cite{minematsu2020visualization} &                                 eye-tracker, mouse &                                                 15 &                                     Neural Network &                          Precision,Recall,F1-score \\ \hline
             \cite{sharma2020eye} &                                eye-tracker, survey &                                                 40 &                     GAM, Gaussian Process, RF, SVM &                  Normalized Root Mean Square Error \\ \hline
     \cite{duraes2019intelligent} &                        eye-tracker, keyboard, logs &                                                 22 &          SVM, kNN, Naive Bayes, Neural Network, RF &                                           Accuracy \\ \hline
        \cite{pham2018predicting} &                                             webcam &                                                 29 &                                                SVM &                               Accuracy, Kappa, AUC \\ \hline
      \cite{sandhu2017evaluation} &                         EEG, eye-tracker, HRV, GSR &                                                 10 &                                   SVM, Rule system &              Accuracy, Precision, Recall, F1-score \\ \hline
            \cite{abe2015towards} &                                             webcam &                                                  2 &                                                kNN &              Accuracy, Precision, Recall, F1-score \\
\bottomrule

\end{tabularx}

\label{tab:biosensors}
\end{table*}

\subsection{Public availability of datasets}

An important aspect in the analysis of MmLA research is the availability of public datasets, which are essential for ensuring reproducibility, enabling benchmarking across studies, and fostering methodological innovation. Among all the reviewed studies, we found that only one \cite{liu2024eeg} that provides access to several publicly available datasets for EEG-based multimodal emotion recognition. These datasets include synchronized physiological signals and annotations, facilitating further research in the field.

However, for the remaining studies in our review, no publicly accessible datasets were identified. In most cases, the datasets used were collected specifically for the individual experiments and have not been released. This represents a notable limitation in the field, as it restricts comparative analysis and slows the advancement of robust, generalizable models in multimodal learning analytics. 

\subsection{Pre-processing and Data Fusion}

30 papers reported information about the algorithms they use. Before using the algorithms, 27 studies report information about pre-processing and fusion of the data from biosensors. 18 studies (66.66\%) perform feature extraction from the signals and apply multimodal data fusion techniques to integrate information from multiple sources. 

One of the most common approaches to integrating data from multiple sources is early fusion, also known as feature-level concatenation. This method involves extracting features independently from each modality and then concatenating these feature vectors into a single, unified representation before feeding them into a machine learning model.
In the reviewed literature, 13 studies out of 18 (72.22\%) applied early or feature-level concatenation strategies to integrate multimodal data after extracting features of each source \cite{abdelmaboud2022machine, duraes2019intelligent, gao2022learning, gupta2024artificial, kawamura2020estimation, kawamura2021detecting, liao2022deploying, nguyen2021multimodal, pham2018predicting, sharma2020eye, van2020facilitating, zhou2024state, zhu2023integrating}. For instance, \cite{nguyen2021multimodal} developed a deep learning model for detecting types of regulatory interactions in collaborative learning by combining features extracted from electrodermal activity (EDA), video, and audio data. Each modality was preprocessed and features were extracted independently. These feature vectors were then concatenated and passed through a dense layer for classification. The results showed that the fused feature set outperformed single-modality models. 

Another approach is decision-level fusion, in which each modality is processed independently, typically using separate models or classifiers, and their individual decisions (e.g., class probabilities or labels) are then combined using strategies such as majority voting, weighted averaging, or rule-based systems. This technique is particularly useful when modalities are heterogeneous or when synchronization at the feature level is impractical. 6 out of 18 (33.33\%) use decision-level fusion \cite{alyuz2017unobtrusive, liao2022deploying, sandhu2017evaluation, zhu2023integrating, kawamura2020estimation, zhang2023evaluation}. For example, in \cite{zhang2023evaluation}, features are first extracted independently from two modalities: facial expression images and textual comments. The extracted features are then classified separately using two distinct models: an AdaBoost classifier for facial expressions and a K-Nearest Neighbors (KNN) classifier for textual data. Finally, the classification outputs are fused at the decision level using a weighted sum rule based on the posterior probabilities of each classifier. In \cite{kawamura2020estimation},  features were independently extracted from two modalities, facial video and seat pressure data, to estimate students’ wakefulness during video-based lectures. From facial video, 45 features were computed, including blink-related indicators and head pose estimations (yaw, pitch, roll). Simultaneously, 80 features were derived from pressure distribution data captured like mean or standard deviation.

The authors compared two multimodal fusion strategies: early fusion, where features from facial video and seat pressure data were concatenated at the feature level, and decision-level fusion, where classification outputs from two separate CatBoost models, one trained on facial features and the other on seat pressure features, were combined. In the decision-level fusion approach, the posterior class probabilities from both classifiers were integrated using a weighted sum formula.

Similarly, in \cite{alyuz2017unobtrusive}, three independent Random Forest classifiers were trained on distinct modalities, appearance (upper-body video), context-performance (interaction logs), and mouse movements, to detect students’ behavioral engagement. The classification outputs were then fused at the decision level using several strategies: majority voting, highest confidence, and a proposed hybrid majority voting (HMV) method where they aggregate the votes from all individual decision trees across the Random Forests of each modality.

An alternative to early and decision-level fusion is intermediate fusion, in which each modality is first processed independently to extract high-level representations—such as embeddings or latent features—before these representations are combined and passed to a joint classifier. Unlike early fusion, where raw or low-level features are concatenated directly, intermediate fusion allows each modality to be transformed into a compact and semantically richer representation using modality-specific neural networks or encoders. These representations are then fused through concatenation, attention mechanisms, or neural fusion layers, and used for downstream classification. Only 1 out of the 18 studies uses this technique \cite{zhu2023integrating}. In particular, the authors proposed a multimodal model to predict students’ academic performance by integrating three types of data: navigation behavior, resource interaction, and self-assessment. Each modality was first processed separately using dedicated encoding subnetworks that learned modality-specific latent representations. These intermediate representations were then concatenated and passed through fully connected layers for final prediction.

Additionally, we also found two review studies that analyzed the use of feature fusion within their respective domains, discussing its implementation, benefits, and challenges.\cite{wang2021towards} analyzes multimodal data integration, including feature fusion techniques, in the context of eye-tracking and learning analytics for building intelligent and collaborative learning environments. Similarly, \cite{liu2024eeg} examines fusion strategies in EEG-based multimodal emotion recognition, highlighting how different levels of fusion contribute to the accuracy and robustness of affective computing systems.

In addition to fusion strategies, several studies also report on data preparation techniques. 7 studies (25.93\%) apply data cleansing \cite{chen2022decade, wang2021towards, van2020facilitating, han2023making, munshi2019personalization, zhu2023integrating, abe2015towards}, and 6 studies (22.22\%) normalize the data \cite{kawamura2021detecting, abdelmaboud2022machine, kawamura2020estimation, di2017learning, duraes2019intelligent, sandhu2017evaluation}. TCN is also used in \cite{zhou2024state}, and a specific toolbox is used in \cite{liu2023evaluation}.

\subsection{Algorithms}

\begin{figure}[h]
    \centering
    \includegraphics[width=\linewidth]{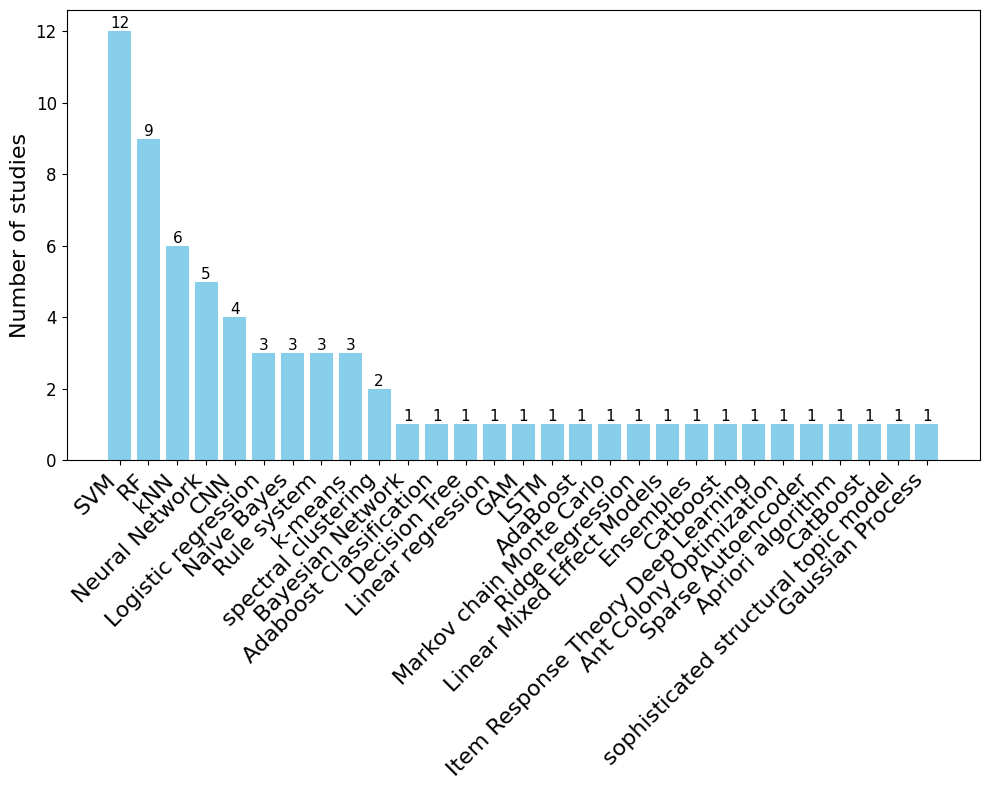}
    \caption{Algorithms used in the different studies}
    \label{fig:algorithms}
\end{figure}

A total of 30 papers reported the algorithms used, with most employing supervised learning (26 studies), while only three used unsupervised methods and one applied temporal analysis.As summarized in Table \ref{tab:biosensors}, most studies tried several algorithms. To determine the prominence of each algorithm, the number of appearances of each algorithm has been plotted in Figure \ref{fig:algorithms}. Support Vector Machine (SVM) is the most used algorithm (12 studies, 40.0\%) \cite{kawamura2021detecting, li2014online, gupta2024artificial, li2024measuring, oder2022automatically, lee2023attention, liao2022deploying, gao2022learning, sharma2020eye, duraes2019intelligent, pham2018predicting, sandhu2017evaluation}, followed by Random Forest (RF) (9 studies, 30.0\%) \cite{chejara2023exploring, kawamura2021detecting, li2014online, azevedo2022lessons, lee2023attention, liao2022deploying, gao2022learning, sharma2020eye, duraes2019intelligent}, k-NN (6 studies, 20.0\%) \cite{li2014online, zhang2023evaluation, lee2023attention, gao2022learning, duraes2019intelligent, abe2015towards}, Neural Network (5 studies, 16.67\%) \cite{liu2024eeg, zhu2023integrating, liao2022deploying, minematsu2020visualization, duraes2019intelligent} and CNN (4 studies, 13.33\%) \cite{nguyen2021multimodal, minematsu2019region, liu2024eeg, zhu2023integrating}. While CNNs were primarily applied to image-based data (e.g., heatmaps, screenshots), the remaining neural network studies did not rely on this specific type of input. Therefore, CNN-based approaches were analyzed separately from other neural network implementations to provide a clearer distinction in methodology.

\subsection{Evaluation Metrics}

Beyond algorithm selection, 28 studies reported details on evaluation metrics used to assess model performance. The most frequently employed metric was accuracy (13 studies, 46.43\%) \cite{zhou2024state, li2014online, abdelmaboud2022machine, gupta2024artificial, oder2022automatically, zhang2023evaluation, liu2024eeg, lee2023attention, gao2022learning, duraes2019intelligent, pham2018predicting, sandhu2017evaluation, abe2015towards}, followed by F1-score (12 studies 42.86\%) \cite{zhou2024state, kawamura2021detecting, gupta2024artificial, alyuz2017unobtrusive, li2024measuring, oder2022automatically, liu2024eeg, zhu2023integrating, gao2022learning, minematsu2020visualization, sandhu2017evaluation, abe2015towards}.
Additionally, recall was reported in 7 studies (25.0\%) \cite{gupta2024artificial, oder2022automatically, liu2024eeg, gao2022learning, minematsu2020visualization, sandhu2017evaluation, abe2015towards}, as was precision, also in 7 studies (25.0\%) \cite{gupta2024artificial, oder2022automatically, liu2024eeg, gao2022learning, minematsu2020visualization, sandhu2017evaluation, abe2015towards}.

However, cross-study comparisons remain challenging due to significant differences in datasets and sensor modalities. Most datasets were collected specifically for each study, limiting the generalizability of reported performance metrics. Furthermore, variations in sensor configurations and data preprocessing techniques introduce additional complexity when interpreting and comparing results across different research works.

\subsection{Monitored Population}
We identified 34 experiments utilizing biometric sensors and multimodal data among the selected papers, although some studies did not report the exact number of students monitored. The median number of participants across these studies was $29.5$, while the mean was $50.82$ ($\text{SD} = 82.35$).

Only one study monitored more than 100 students. In that study \cite{gupta2024artificial}, the authors collected data from 500 students using EEG bands and webcams to compute an attention index. The EEG signals were analyzed alongside facial cues captured from webcams, including eye blinks, head movements, and facial expressions, to classify students' cognitive states as attentive or inattentive.

The other studies with more participants only utilized a few sensors: in \cite{zhang2023evaluation}, 100 students were monitored using a webcam to extract facial expressions and detect emotions; in \cite{munshi2019personalization}, only data from logs and surveys were considered to track students in self-regulated learning; in \cite{nguyen2021multimodal}, EDA sensors, webcams, and audio were captured from 94 students for detecting types of interactions in collaborative learning; and in \cite{oder2022automatically}, 78 participants were monitored using an eye-tracker to detect familiarity with a website.

Regarding the average age of the students in the studies, only 18 experiments reported this information. All studies monitored teenagers and young adults: 5 studies (27.78\%) monitored teenagers between 14 and 18 years old \cite{chejara2023exploring, munshi2019personalization, alyuz2017unobtrusive, gao2022learning, duraes2019intelligent}, 9 studies (50.0\%) monitored students with average age between 18 and 24 years old \cite{liu2023evaluation, kawamura2021detecting, kawamura2020estimation, minematsu2019region, li2024measuring, zhu2023integrating, lee2023attention, ciolacu2021education, sandhu2017evaluation}, and 4 studies (22.22\%) monitored students with average age between 25 and 30 years old \cite{zhou2024state, di2017learning, liao2022deploying, pham2018predicting}. Additionally, the educational level of the students was reported in 19 experiments, the majority were university students (10 studies, 52.63\%) \cite{zhou2024state, kawamura2021detecting, kawamura2020estimation, minematsu2019region, jia2021design, montebello2018assisting, zhu2023integrating, lee2023attention, minematsu2020visualization, sharma2020eye} or graduates students (4 studies, 21.05\%) \cite{liu2023evaluation, li2024measuring, di2017learning, liao2022deploying}. There are also 3 studies (15.79\%) \cite{chejara2023exploring, munshi2019personalization, duraes2019intelligent} monitoring high school students and 2 studies (10.53\%) \cite{alyuz2017unobtrusive, gao2022learning} monitoring school students.

\section{RQ3. Limitations and future work}\label{ss:future work}
Understanding and predicting students' behavior is challenging and comes with certain limitations. Upon analyzing the studies, most agreed on the need to enhance models by collecting more data through monitoring a larger number of students and analyzing additional features \cite{mu2023bcrnet, praharaj2021literature, toala2019human, liao2022deploying, duraes2019intelligent, minematsu2020visualization, sandhu2017evaluation, pham2018predicting, bradavc2022design, sharma2020eye} or by incorporating additional sensors \cite{jia2021design, abe2015towards}. 

For instance, in \cite{pham2018predicting}, a multimodal intelligent tutor designed to enhance MOOC learning experiences by detecting students' emotions using mobile devices was tested with only 29 participants. To extend its applicability beyond the laboratory, the authors plan to conduct large-scale, longitudinal studies in real-world learning environments, enabling a more comprehensive evaluation of its effectiveness in everyday student interactions. Similarly, in \cite{minematsu2020visualization}, a system was developed to support teachers by analyzing student learning behaviors using clickstream data and eye movement tracking. However, due to the limitations of eye-tracking, which requires specialized hardware and is difficult to scale beyond small groups, the study was conducted with only 15 students under controlled laboratory conditions.

Another challenge highlighted in \cite{liu2024eeg}, is the need for improved data fusion techniques, particularly when integrating signals from multiple sensors. The authors emphasize that physiological signals exhibit variable sensitivity to external noise, complicating their synchronization and reducing model accuracy. Additionally, they underscore the limitations of working with missing values, stressing the need for robust methodologies and effective data imputation techniques.

Cost and environmental factors also play a crucial role in data collection. In \cite{zhu2023integrating}, the authors advocate for the use of more affordable sensors while highlighting concerns about the artificiality of controlled environments, which can lead participants to behave differently than they would in real-world settings. As a solution, they propose gathering data in more natural learning environments. However, they also point out that the lack of standardized benchmarking and publicly available code hinders progress and reproducibility in the field.

Several studies explore new methods for improving learning experiences. In \cite{sharma2020eye}, future work includes using gaze analytics to identify student profiles and provide gaze-aware feedback to enhance learning outcomes. Similarly, in \cite{lee2023attention}, they proposed using different reader profiles to generate feedback for online reading. Feedback using EEG band data is also pointed out as future work in \cite{han2023multimodal}, as neurofeedback is currently only applied in fields other than online learning. 

Others studies only proposed a theoretical design, intending to implement it in the future \cite{huang2017research, wiedbusch2021theoretical, nguyen2021multimodal, minematsu2019region}, while others proposed improvements to their initial approaches. For instance, \cite{gao2022learning} proposed enhancing labeling strategies by integrating additional multimodal data, including students’ eye trajectory, head posture, and facial expressions, to achieve more accurate attention level classification. Additionally, they suggested incorporating expert annotations alongside EEG- and multimodal-based labels, as well as self-reported data, to establish a more reliable three-way labeling system.

\cite{coltey2021generalized} proposed facilitating the implementation of virtual reality learning environments by developing a standardized framework that enables dynamic adaptation of immersive content in real time. This included refining a generalizable learning process description language, improving AI-driven adaptive learning systems, and integrating multimodal data collection to personalize learning experiences. Additionally, they proposed optimizing performance for lower-cost virtual reality devices and automating the conversion of traditional educational content into immersive learning environments.

\cite{cloude2022system} emphasized the importance of developing systems such as intelligent tutors and AI models capable of analyzing captured data in real time. Similarly, \cite{pham2018predicting, bradavc2022design} highlighted the need for these systems to be compatible with platforms like Moodle and edX to ensure accessibility, enabling seamless integration with existing educational infrastructures and facilitating widespread adoption across diverse learning environments. Additionally, \cite{duraes2019intelligent} stressed the importance of personalizing these systems to adapt to individual student needs, for example, by defining different user profiles and leveraging AI-driven insights to provide tailored learning experiences that enhance engagement and improve learning outcomes.

In the case of \cite{caballe2015towards}, after proposing an approach for detecting emotions in online learning, the authors aim to overcome the current limitations of online platform through affective feedback tools and emotional assessment mechanisms. Their goal is to enhance personalized learning, increase motivation, and reduce dropout rates. Additionally, they foresee that integrating emotion recognition technologies and artificial intelligence models will enable adaptive teaching strategies and foster greater social interaction in digital learning environments, ultimately optimizing the educational experience.

Some studies proposed focusing on new variables for prediction, such as dropout \cite{zhou2024state} or teachers' attention \cite{kawamura2021detecting}. Others proposed combining techniques from other fields, such as pedagogy and human-computer interaction theories \cite{wang2021towards}, or natural language processing techniques that could help students in self-regulated learning \cite{azevedo2022lessons}.

\section{Discussion}

The findings of this systematic review highlight that the integration of biosensors and MmLA into online learning environments remains in an initial phase, predominantly characterized by feasibility studies and controlled experimental settings. While the diversity of sensors and methodological approaches suggests a dynamic and rapidly evolving research landscape, it also reveals a lack of standardization, which hinders cross-study comparison and limits reproducibility.

A major limitation across studies is the small and homogeneous sample sizes, which significantly affect the generalizability of results. Additionally, the near absence of publicly available datasets restricts the advancement of robust, comparative, and replicable models. This methodological gap underscores the need for open protocols for data sharing and more representative, longitudinal studies to validate findings across diverse populations. Recent efforts have emerged to address this gap. For example, \cite{juvrik2025experimental} released an openly accessible dataset collected from 110 university students using eye-tracking to study the impact of metacognitive prompts during self-regulated multimedia learning tasks in a controlled laboratory environment. Similarly, the IMPROVE dataset \cite{daza2024improve} further support this trend by providing a publicly available multimodal dataset that captures behavioral, biometric, and physiological data from 120 university students to study the impact of mobile phone usage during online learning sessions.

The most commonly used sensors, such as webcams, eye-trackers, and EEG bands, demonstrate considerable potential for detecting cognitive and emotional states such as attention, motivation, and academic performance. However, their large-scale deployment still faces logistical, financial, and ethical challenges, particularly in terms of privacy, data sensitivity, and student acceptance \cite{nguyen2023ethical, zhu2023integrating}.

From a methodological perspective, the prevalent reliance on Support Vector Machines (SVM) and Random Forests reflects a tendency toward classical machine learning algorithms that are relatively robust in low-data scenarios. This preference is closely linked to the limited availability of large, annotated datasets in the field, which poses significant challenges for training more complex, data-intensive models such as deep neural networks. Although there is an emerging interest in the application of convolutional neural networks (CNNs) for analyzing visual and physiological signals, their deployment remains constrained by insufficient data volume and variability. Moreover, the overall performance of multimodal systems continues to be affected by heterogeneity in data fusion strategies. While feature-level (early) fusion remains the most commonly implemented technique due to its straightforward integration of multiple modalities, intermediate and decision-level fusion approaches are comparatively underexplored, despite their potential to enhance model robustness and interpretability in multimodal learning contexts.

Another recurring theme is the gap between experimental setups and real-world applicability. Several studies propose future work focused on implementing intelligent tutoring systems and adaptive platforms that can integrate with mainstream learning management systems (e.g., Moodle, edX). However, the pedagogical effectiveness of these interventions remains largely untested. There is a clear need for empirical studies that evaluate whether multimodal feedback or data-based interventions genuinely improve learning outcomes, engagement, or retention. For example, \cite{nguyen2025guidelines} highlights that while recent advances in GenAI enable more personalised and multimodal feedback, their impact on learning remains under-explored. Rigorous research is needed to evaluate the pedagogical effectiveness of GenAI-driven feedback, particularly regarding its emotional impact, potential for overreliance, and alignment with higher-order learning goals.

Finally, it is notable that many studies operate with a strong technological focus but lack theoretical grounding. Few explicitly connect their analytics frameworks with learning theories. Bridging this gap is essential for ensuring that multimodal data is interpreted meaningfully in educational contexts. The field would benefit greatly from interdisciplinary collaborations that combine expertise in learning sciences, artificial intelligence, and human-computer interaction.

In summary, while the potential of biosensor-integrated MmLA for enhancing personalized and adaptive online learning is promising, current research remains largely exploratory. Advancing in this field will require not only methodological rigor and standardization but also a more explicit engagement with pedagogical goals and learning theories.

\section{Conclusions}
In this systematic review, we examined 54 academic papers to analyze the technologies and algorithms being utilized in online learning to understand and predict student behavior through biosensors and multimodal data. By synthesizing these diverse studies, we have provided an overview of key aspects including methods for detecting and predicting student behaviors, approaches for capturing and processing physiological data, and identified limitations and future directions in this emerging field.

By reviewing the selected studies, we address three key research questions. First, we examine how researchers have detected and predicted students' behaviors using Multimodal Learning Analytics and biosensors in online learning contexts, and how they have evaluated learning presence and student motivation. Second, we explore how physiological data from students has been captured and processed, how algorithms have been applied to these signals, and how study populations have been distributed. Finally, we identify the main limitations reported in the literature and the proposed directions for future research.

Despite the diversity of the studies, we have been able to extract insights by classifying them and identifying common points and we have provided an overview of the different behaviors and biosensors used in the field. This allowed us to highlight both well-established areas of multimodal research as well as gaps and opportunities for further investigation. The prevalence of ``viability studies'' (approaches designed to demonstrate practical approaches) reflects the relative youth of this field, with much exploratory work still needed.

Our analysis revealed certain trends, such as the widespread use of webcams, eye-trackers, and EEG bands for behavior detection and prediction. We also found Support Vector Machines and Random Forests to be the dominant modeling approaches, with accuracy as the primary evaluation metric. These patterns raise important questions about whether more focused studies on specific sensor types are needed, and if the field should expand its methodological toolkit.

A key limitation identified across studies was the relatively small sample sizes, with most monitoring only a limited number of students. The lack of common, standardized datasets also poses challenges for comparing results across studies. Additionally, we found the majority of research focused on teenagers and young adults, indicating a need for future work to include a broader age range to improve generalizability and reduce potential age bias.

In conclusion, while the integration of biosensors and multimodal learning analytics holds promise for understanding student behavior and enabling adaptive, personalized online learning, the evidence from the reviewed studies suggests that this potential remains largely theoretical. The majority of current work focuses on feasibility studies with small, homogeneous participant groups and highly controlled environments, raising concerns about generalizability and validity. Future research must rigorously interrogate when, how and for whom multimodal data actually improves learning outcomes. This includes comparative studies that benchmark biosensor-based interventions against less invasive or more scalable alternatives. Moreover, addressing limitations around data availability, methodological diversity, standardizing data practices, and diversifying demographic representation is crucial for advancing from a phase of technological experimentation to one of substantive, pedagogically grounded innovation. As online learning continues to grow in importance, further research in this area holds significant potential to enhance the ability to design adaptive and personalized learning experiences. However, without this shift toward methodological rigor and inclusive, scalable research design, the field risks prioritizing technological complexity over real, measurable educational impact.

\section*{Acknowledgments}
Support by projects: Cátedra ENIA UAM-VERIDAS en IA Responsable (NextGenerationEU PRTR TSI-100927-2023-2), HumanCAIC (TED2021-131787B-I00 MICINN) and SNOLA (RED2022-134284-T)

%%
%% The next two lines define the bibliography style to be used, and
%% the bibliography file.
\bibliographystyle{IEEEtran}
\bibliography{biblio}

\end{document}